# Analysis of Prime Reciprocal Sequences in Base 10

Sumanth Kumar Reddy Gangasani

## Introduction

The motivation for this study is the determination of structural redundancy in the prime reciprocals in base 10 in a manner that parallels a similar study for binary prime reciprocals [1-3]. Prime reciprocal sequences (also called d-sequences or decimal sequences for base 10) have applications in coding and cryptography [4-10] and for generation of random sequences [11-13].

We show that there are several different kinds of structural relationships amongst the digits in reciprocal sequences which are best classified with respect to the digit in the least significant place of the prime. Also, the frequency of digit 0 exceeds that of every other digit when the entire set of prime reciprocal sequences is considered.

## Generating the Sequence

The formula used to generate the digits of the prime reciprocal sequences is as below [6]:

$$a_i = (l(10^i \bmod p)) \bmod 10$$

where $a_i$ is the $i$th digit in the sequence, $p$ is the prime number and the value of $l$ is given as below

$$\begin{aligned}
l &= 1 & \text{if,} \quad p \bmod 10 &= 9 \\
l &= 3 & \text{if,} \quad p \bmod 10 &= 3 \\
l &= 7 & \text{if,} \quad p \bmod 10 &= 7 \\
l &= 9 & \text{if,} \quad p \bmod 10 &= 1
\end{aligned}$$

## Digit Frequencies

The least significant digit of the prime can be 1, 3, 7 or 9. One would expect different results for each of these cases. Furthermore, for any of these there would be further structure given depending on what the second least significant digit is when the sequence is half length.

Our analysis shows that Kak's conjecture[6] that there are more 0s than 1s in most binary prime reciprocal sequences holds not only for binary and ternary prime reciprocal sequences[14] but also for decimal prime reciprocal sequences.

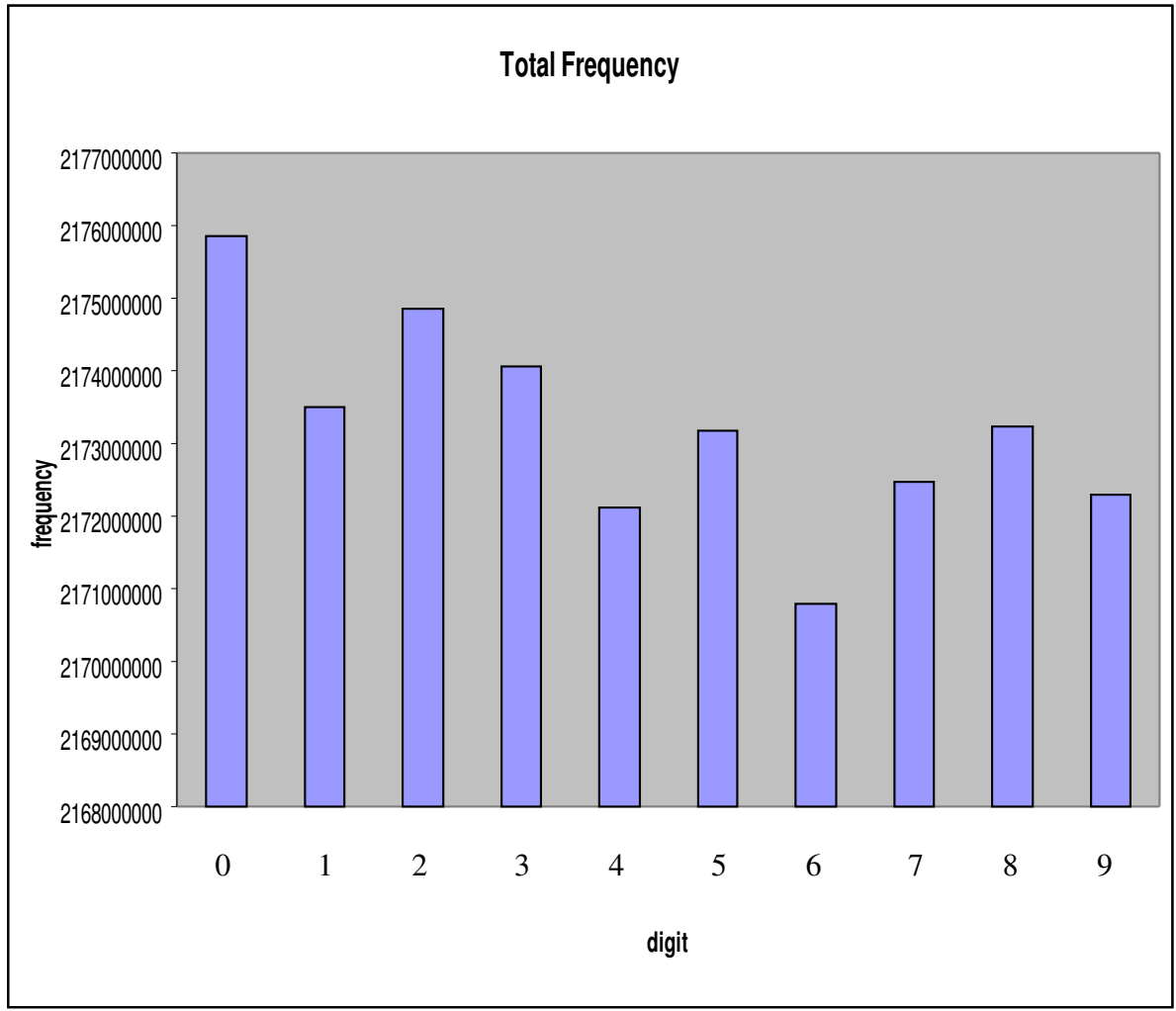

**Figure 1.** Frequencies of digits 0 through 9 for prime reciprocal sequences

The results of Figure 1 are based on prime reciprocal sequences for primes up to 999983. We expect that the excess of 0s over the other digits will persist as the values of primes are increased further.

## Structure in the Decimal Reciprocal Sequences

The structure in the decimal expansion of various prime reciprocals becomes clear when we consider the primes with specific ending. The four cases, with the endings of 1, 3, 7, and 9, are given below.

## Sequence when primes end in 1:

Full length sequences ending in 1 are observed to have equal number of each digit in every sequence.

In any half length sequence ending in 1, when the second most significant digit is even, one would see that the frequency of 0 is equal to the frequency of 9 in every sequence, and maximum in many of the sequences. Digits 1, 2, 4, 5, 7 and 8 have the same frequency in each sequence and have either maximum or minimum frequency. The frequency of 3 is also seen to be the same as the frequency of 6 in every such sequence, and minimum in some sequences where 0s and 9s do not have maximum frequency.

**Table 1:** Frequency distribution in the prime reciprocal sequences to the base-10 when the primes end in 1 and the 2$^{nd}$ least significant digit is even.

| Prime No | 0s | 1s | 2s | 3s | 4s | 5s | 6s | 7s | 8s | 9s |
|---|---|---|---|---|---|---|---|---|---|---|
| 601 | 35 | 28 | 28 | 31 | 28 | 28 | 31 | 28 | 28 | 35 |
| 3001 | 164 | 146 | 146 | 148 | 146 | 146 | 148 | 146 | 146 | 164 |
| 84401 | 4231 | 4251 | 4251 | 4116 | 4251 | 4251 | 4116 | 4251 | 4251 | 4231 |
| 473201 | 23737 | 23692 | 23692 | 23487 | 23692 | 23692 | 23487 | 23692 | 23692 | 23737 |
| 965801 | 48395 | 48350 | 48350 | 48005 | 48350 | 48350 | 48005 | 48350 | 48350 | 48395 |
| 6121 | 325 | 300 | 300 | 305 | 300 | 300 | 305 | 300 | 300 | 325 |
| 17321 | 881 | 869 | 869 | 842 | 869 | 869 | 842 | 869 | 869 | 881 |
| 317921 | 15925 | 15933 | 15933 | 15756 | 15933 | 15933 | 15756 | 15933 | 15933 | 15925 |
| 342521 | 17177 | 17164 | 17164 | 16961 | 17164 | 17164 | 16961 | 17164 | 17164 | 17177 |
| 940721 | 47144 | 47079 | 47079 | 46799 | 47079 | 47079 | 46799 | 47079 | 47079 | 47144 |
| 5441 | 276 | 277 | 277 | 253 | 277 | 277 | 253 | 277 | 277 | 276 |
| 22441 | 1157 | 1107 | 1107 | 1132 | 1107 | 1107 | 1132 | 1107 | 1107 | 1157 |
| 166841 | 8405 | 8337 | 8337 | 8294 | 8337 | 8337 | 8294 | 8337 | 8337 | 8405 |
| 394241 | 19795 | 19736 | 19736 | 19557 | 19736 | 19736 | 19557 | 19736 | 19736 | 19795 |
| 924641 | 46345 | 46269 | 46269 | 46008 | 46269 | 46269 | 46008 | 46269 | 46269 | 46345 |
| 761 | 40 | 39 | 39 | 33 | 39 | 39 | 33 | 39 | 39 | 40 |
| 73361 | 3687 | 3683 | 3683 | 3604 | 3683 | 3683 | 3604 | 3683 | 3683 | 3687 |
| 104761 | 5296 | 5217 | 5217 | 5243 | 5217 | 5217 | 5243 | 5217 | 5217 | 5296 |
| 371561 | 18605 | 18626 | 18626 | 18407 | 18626 | 18626 | 18407 | 18626 | 18626 | 18605 |
| 899161 | 45102 | 44920 | 44920 | 44928 | 44920 | 44920 | 44928 | 44920 | 44920 | 45102 |
| 881 | 49 | 44 | 44 | 39 | 44 | 44 | 39 | 44 | 44 | 49 |
| 5281 | 282 | 257 | 257 | 267 | 257 | 257 | 267 | 257 | 257 | 282 |
| 42281 | 2117 | 2133 | 2133 | 2054 | 2133 | 2133 | 2054 | 2133 | 2133 | 2117 |
| 309481 | 15561 | 15449 | 15449 | 15462 | 15449 | 15449 | 15462 | 15449 | 15449 | 15561 |
| 989081 | 49519 | 49508 | 49508 | 49227 | 49508 | 49508 | 49227 | 49508 | 49508 | 49519 |

In any half length sequence ending in 1, when the second most significant digit is odd, one would see that the sum of the frequencies of complementary digits is equal to the integral part of *p*/10. Number of 1s, 5s and 6s are equal and the number of 3s, 4s and 8s are also equal. It is also seen that either 0s or 2s have the maximum frequency and consequently either 7s or 9s have minimum frequency.

Table 2: Frequency distribution in the prime reciprocal sequences to the base-10 when the primes end in 1 and the 2nd least significant digit is odd.

| Prime No | 0s | 1s | 2s | 3s | 4s | 5s | 6s | 7s | 8s | 9s |
|---|---|---|---|---|---|---|---|---|---|---|
| 911 | 58 | 47 | 50 | 44 | 44 | 47 | 47 | 41 | 44 | 33 |
| 3511 | 185 | 181 | 192 | 170 | 170 | 181 | 181 | 159 | 170 | 166 |
| 33311 | 1771 | 1671 | 1682 | 1660 | 1660 | 1671 | 1671 | 1649 | 1660 | 1560 |
| 388111 | 19510 | 19452 | 19545 | 19359 | 19359 | 19452 | 19452 | 19266 | 19359 | 19301 |
| 997511 | 50165 | 49925 | 50024 | 49826 | 49826 | 49925 | 49925 | 49727 | 49826 | 49586 |
| 631 | 33 | 34 | 39 | 29 | 29 | 34 | 34 | 24 | 29 | 30 |
| 5231 | 290 | 266 | 275 | 257 | 257 | 266 | 266 | 248 | 257 | 233 |
| 77431 | 3905 | 3904 | 3969 | 3839 | 3839 | 3904 | 3904 | 3774 | 3839 | 3838 |
| 454031 | 22950 | 22738 | 22811 | 22665 | 22665 | 22738 | 22738 | 22592 | 22665 | 22453 |
| 911831 | 46114 | 45646 | 45755 | 45537 | 45537 | 45646 | 45646 | 45428 | 45537 | 45069 |
| 151 | 8 | 9 | 12 | 6 | 6 | 9 | 9 | 3 | 6 | 7 |
| 2351 | 144 | 120 | 125 | 115 | 115 | 120 | 120 | 110 | 115 | 91 |
| 57751 | 2909 | 2911 | 2958 | 2864 | 2864 | 2911 | 2911 | 2817 | 2864 | 2866 |
| 288551 | 14666 | 14459 | 14522 | 14396 | 14396 | 14459 | 14459 | 14333 | 14396 | 14189 |
| 998951 | 50399 | 49997 | 50096 | 49898 | 49898 | 49997 | 49997 | 49799 | 49898 | 49496 |
| 1471 | 76 | 78 | 87 | 69 | 69 | 78 | 78 | 60 | 69 | 71 |
| 23071 | 1185 | 1164 | 1185 | 1143 | 1143 | 1164 | 1164 | 1122 | 1143 | 1122 |
| 76871 | 3934 | 3856 | 3881 | 3831 | 3831 | 3856 | 3856 | 3806 | 3831 | 3753 |
| 597671 | 30285 | 29918 | 29987 | 29849 | 29849 | 29918 | 29918 | 29780 | 29849 | 29482 |
| 996271 | 49978 | 49891 | 50046 | 49736 | 49736 | 49891 | 49891 | 49581 | 49736 | 49649 |
| 991 | 51 | 53 | 60 | 46 | 46 | 53 | 53 | 39 | 46 | 48 |
| 9391 | 472 | 482 | 507 | 457 | 457 | 482 | 482 | 432 | 457 | 467 |
| 27791 | 1476 | 1397 | 1412 | 1382 | 1382 | 1397 | 1397 | 1367 | 1382 | 1303 |
| 347591 | 17826 | 17397 | 17432 | 17362 | 17362 | 17397 | 17397 | 17327 | 17362 | 16933 |
| 878191 | 44069 | 43998 | 44175 | 43821 | 43821 | 43998 | 43998 | 43644 | 43821 | 43750 |

**Sequence when primes end in 3:**

In full length sequences, number of 3s and 6s is 1 more than the number of any other digit and all other digits have the same frequency.

In a half length sequence that ends with 3, if the 2nd least significant digit is even, the sum of the frequencies of the complementary digits is equal to the integral part of *p*/10. Number of 0s and 9s is always seen to be equal. Digits 1, 4 and 7 are seen to be equally frequent and since the complementary digits are 8, 5 and 2 respectively, number of 2s, 5s and 8s is also equal. The digit 3 has maximum frequency and as expected the complementary digit 6 has minimum frequency.

**Table 3:** Frequency distribution in the prime reciprocal sequences to the base-10 when the prime ends in 3 and the 2nd least significant digit is even.

| Prime No | 0s | 1s | 2s | 3s | 4s | 5s | 6s | 7s | 8s | 9s |
|---|---|---|---|---|---|---|---|---|---|---|
| 2203 | 110 | 101 | 119 | 127 | 101 | 119 | 94 | 101 | 119 | 110 |
| 5003 | 250 | 251 | 249 | 272 | 251 | 249 | 229 | 251 | 249 | 250 |
| 64403 | 3220 | 3245 | 3195 | 3308 | 3245 | 3195 | 3133 | 3245 | 3195 | 3220 |
| 431603 | 21580 | 21593 | 21567 | 21869 | 21593 | 21567 | 21292 | 21593 | 21567 | 21580 |
| 996803 | 49840 | 49936 | 49744 | 50145 | 49936 | 49744 | 49536 | 49936 | 49744 | 49840 |
| 523 | 26 | 24 | 28 | 36 | 24 | 28 | 17 | 24 | 28 | 26 |
| 5923 | 296 | 275 | 317 | 328 | 275 | 317 | 265 | 275 | 317 | 296 |
| 92723 | 4636 | 4628 | 4644 | 4733 | 4628 | 4644 | 4540 | 4628 | 4644 | 4636 |
| 354323 | 17716 | 17718 | 17714 | 17932 | 17718 | 17714 | 17501 | 17718 | 17714 | 17716 |
| 954323 | 47716 | 47810 | 47622 | 48110 | 47810 | 47622 | 47323 | 47810 | 47622 | 47716 |
| 443 | 22 | 23 | 21 | 29 | 23 | 21 | 16 | 23 | 21 | 22 |
| 7643 | 382 | 396 | 368 | 412 | 396 | 368 | 353 | 396 | 368 | 382 |
| 49043 | 2452 | 2453 | 2451 | 2522 | 2453 | 2451 | 2383 | 2453 | 2451 | 2452 |
| 382843 | 19142 | 19025 | 19259 | 19387 | 19025 | 19259 | 18898 | 19025 | 19259 | 19142 |
| 844243 | 42212 | 42025 | 42399 | 42542 | 42025 | 42399 | 41883 | 42025 | 42399 | 42212 |
| 563 | 28 | 32 | 24 | 38 | 32 | 24 | 19 | 32 | 24 | 28 |
| 8963 | 448 | 460 | 436 | 480 | 460 | 436 | 417 | 460 | 436 | 448 |
| 19763 | 988 | 996 | 980 | 1042 | 996 | 980 | 935 | 996 | 980 | 988 |
| 498163 | 24908 | 24711 | 25105 | 25239 | 24711 | 25105 | 24578 | 24711 | 25105 | 24908 |
| 950363 | 47518 | 47467 | 47569 | 47910 | 47467 | 47569 | 47127 | 47467 | 47569 | 47518 |
| 683 | 34 | 33 | 35 | 43 | 33 | 35 | 26 | 33 | 35 | 34 |
| 6883 | 344 | 323 | 365 | 379 | 323 | 365 | 310 | 323 | 365 | 344 |
| 27283 | 1364 | 1332 | 1396 | 1431 | 1332 | 1396 | 1298 | 1332 | 1396 | 1364 |
| 233083 | 11654 | 11537 | 11771 | 11848 | 11537 | 11771 | 11461 | 11537 | 11771 | 11654 |
| 985483 | 49274 | 48964 | 49584 | 49721 | 48964 | 49584 | 48828 | 48964 | 49584 | 49274 |

In a half length sequence that ends with 3, if the second least significant digit is odd, the complementary digits have equal frequencies. In most of the cases 1s and 8s or 2s and 7s are observed to have maximum frequency. In all these sequences, 3s and 6s or 4s and 5s have the minimum frequency. In other sequences, 0s and 9s have maximum frequency and 1s and 8s have the minimum frequency.

**Table 4:** Frequency distribution in the prime reciprocal sequences to the base-10 when the primes end in 3 and the 2$^{nd}$ least significant digit is odd.

| Prime No | 0s | 1s | 2s | 3s | 4s | 5s | 6s | 7s | 8s | 9s |
|---|---|---|---|---|---|---|---|---|---|---|
| 5413 | 278 | 267 | 282 | 267 | 259 | 259 | 267 | 282 | 267 | 278 |
| 49613 | 2467 | 2522 | 2495 | 2453 | 2466 | 2466 | 2453 | 2495 | 2522 | 2467 |
| 89213 | 4458 | 4498 | 4493 | 4426 | 4428 | 4428 | 4426 | 4493 | 4498 | 4458 |
| 235013 | 11723 | 11855 | 11800 | 11674 | 11701 | 11701 | 11674 | 11800 | 11855 | 11723 |
| 914813 | 45679 | 45939 | 45816 | 45604 | 45665 | 45665 | 45604 | 45816 | 45939 | 45679 |
| 3533 | 178 | 180 | 183 | 172 | 170 | 170 | 172 | 183 | 180 | 178 |
| 18133 | 925 | 890 | 927 | 905 | 886 | 886 | 905 | 927 | 890 | 925 |
| 48733 | 2474 | 2398 | 2473 | 2438 | 2400 | 2400 | 2438 | 2473 | 2398 | 2474 |
| 266333 | 13286 | 13422 | 13361 | 13242 | 13272 | 13272 | 13242 | 13361 | 13422 | 13286 |
| 986933 | 49373 | 49390 | 49443 | 49277 | 49250 | 49250 | 49277 | 49443 | 49390 | 49373 |
| 653 | 32 | 36 | 35 | 30 | 30 | 30 | 30 | 35 | 36 | 32 |
| 6053 | 302 | 312 | 311 | 294 | 294 | 294 | 294 | 311 | 312 | 302 |
| 61253 | 3071 | 3078 | 3095 | 3039 | 3030 | 3030 | 3039 | 3095 | 3078 | 3071 |
| 391453 | 19673 | 19468 | 19669 | 19577 | 19476 | 19476 | 19577 | 19669 | 19468 | 19673 |
| 984853 | 49348 | 49173 | 49384 | 49207 | 49101 | 49101 | 49207 | 49384 | 49173 | 49348 |
| 373 | 20 | 18 | 21 | 18 | 16 | 16 | 18 | 21 | 18 | 20 |
| 6173 | 309 | 315 | 316 | 302 | 301 | 301 | 302 | 316 | 315 | 309 |
| 70573 | 3562 | 3490 | 3557 | 3534 | 3500 | 3500 | 3534 | 3557 | 3490 | 3562 |
| 228773 | 11444 | 11479 | 11490 | 11393 | 11387 | 11387 | 11393 | 11490 | 11479 | 11444 |
| 982973 | 49149 | 49260 | 49261 | 49037 | 49036 | 49036 | 49037 | 49261 | 49260 | 49149 |
| 293 | 14 | 17 | 16 | 13 | 13 | 13 | 13 | 16 | 17 | 14 |
| 8093 | 396 | 427 | 410 | 391 | 399 | 399 | 391 | 410 | 427 | 396 |
| 33493 | 1703 | 1650 | 1707 | 1671 | 1642 | 1642 | 1671 | 1707 | 1650 | 1703 |
| 449693 | 22479 | 22563 | 22552 | 22412 | 22417 | 22417 | 22412 | 22552 | 22563 | 22479 |
| 993893 | 49724 | 49748 | 49807 | 49612 | 49582 | 49582 | 49612 | 49807 | 49748 | 49724 |

**Sequence when primes end in 7:**

In full length sequences, number of 0s, 3s, 6s and 9s is 1 less than the number of other digits.

In a half length sequence that ends with 7, if the second least significant digit is even, sum of the frequencies of complementary digits for 0s and 3s is equal to integral part of $p/10$ and 1 more than integral part of $p/10$ for others. The digits 1, 4 and 7 have equal frequencies and hence the complementary digits 2, 5 and 8 have the same frequency. It is also seen that 3 occurs maximum number of times in every sequence and 6 occurs minimum number of times in every such sequence.

**Table 5:** Frequency distribution in the prime reciprocal sequences to the base-10 when the primes end in 7 and the 2nd least significant digit is even.

| Prime No | 0s | 1s | 2s | 3s | 4s | 5s | 6s | 7s | 8s | 9s |
|---|---|---|---|---|---|---|---|---|---|---|
| 307 | 15 | 13 | 18 | 22 | 13 | 18 | 8 | 13 | 18 | 15 |
| 5507 | 275 | 285 | 266 | 300 | 285 | 266 | 250 | 285 | 266 | 275 |
| 17107 | 855 | 835 | 876 | 907 | 835 | 876 | 803 | 835 | 876 | 855 |
| 195907 | 9795 | 9754 | 9837 | 9952 | 9754 | 9837 | 9638 | 9754 | 9837 | 9795 |
| 953707 | 47685 | 47516 | 47855 | 48132 | 47516 | 47855 | 47238 | 47516 | 47855 | 47685 |
| 827 | 41 | 41 | 42 | 52 | 41 | 42 | 30 | 41 | 42 | 41 |
| 2027 | 101 | 103 | 100 | 116 | 103 | 100 | 86 | 103 | 100 | 101 |
| 23227 | 1161 | 1132 | 1191 | 1222 | 1132 | 1191 | 1100 | 1132 | 1191 | 1161 |
| 139627 | 6981 | 6929 | 7034 | 7107 | 6929 | 7034 | 6855 | 6929 | 7034 | 6981 |
| 963427 | 48171 | 48101 | 48242 | 48459 | 48101 | 48242 | 47883 | 48101 | 48242 | 48171 |
| 947 | 47 | 46 | 49 | 56 | 46 | 49 | 38 | 46 | 49 | 47 |
| 5147 | 257 | 262 | 253 | 281 | 262 | 253 | 233 | 262 | 253 | 257 |
| 33547 | 1677 | 1637 | 1718 | 1758 | 1637 | 1718 | 1596 | 1637 | 1718 | 1677 |
| 197347 | 9867 | 9820 | 9915 | 10027 | 9820 | 9915 | 9707 | 9820 | 9915 | 9867 |
| 995747 | 49787 | 49828 | 49747 | 50087 | 49828 | 49747 | 49487 | 49828 | 49747 | 49787 |
| 467 | 23 | 27 | 20 | 30 | 27 | 20 | 16 | 27 | 20 | 23 |
| 3067 | 153 | 141 | 166 | 176 | 141 | 166 | 130 | 141 | 166 | 153 |
| 25667 | 1283 | 1300 | 1267 | 1334 | 1300 | 1267 | 1232 | 1300 | 1267 | 1283 |
| 313267 | 15663 | 15627 | 15700 | 15839 | 15627 | 15700 | 15487 | 15627 | 15700 | 15663 |
| 992867 | 49643 | 49652 | 49635 | 50035 | 49652 | 49635 | 49251 | 49652 | 49635 | 49643 |
| 787 | 39 | 35 | 44 | 51 | 35 | 44 | 27 | 35 | 44 | 39 |
| 5387 | 269 | 272 | 267 | 301 | 272 | 267 | 237 | 272 | 267 | 269 |
| 16187 | 809 | 819 | 800 | 861 | 819 | 800 | 757 | 819 | 800 | 809 |
| 330587 | 16529 | 16564 | 16495 | 16706 | 16564 | 16495 | 16352 | 16564 | 16495 | 16529 |
| 995987 | 49799 | 49878 | 49721 | 50076 | 49878 | 49721 | 49522 | 49878 | 49721 | 49799 |

In a half length sequences that end in 7, if the second least significant digit is odd, the complementary digits have equal frequencies. In most of the cases 1s and 8s or 2s and 7s are observed to have maximum frequency. In all these sequences, 3s and 6s or 4s and 5s have the minimum frequency. In other sequences, 0s and 9s have maximum frequency and 1s and 8s have the minimum frequency.

This structure of the digit frequencies of half length sequences the end in 7, when compared to the structure of the digit frequencies of half length sequences that end with the complementary digit 3, shows a lot of similarity.

**Table 6:** Frequency distribution in the prime reciprocal sequences to the base-10 when the primes end in 7 and the 2nd least significant digit is odd.

| Prime No | 0s | 1s | 2s | 3s | 4s | 5s | 6s | 7s | 8s | 9s |
|---|---|---|---|---|---|---|---|---|---|---|
| 2917 | 152 | 140 | 153 | 145 | 139 | 139 | 145 | 153 | 140 | 152 |
| 14717 | 739 | 742 | 749 | 726 | 723 | 723 | 726 | 749 | 742 | 739 |
| 74317 | 3769 | 3658 | 3765 | 3720 | 3667 | 3667 | 3720 | 3765 | 3658 | 3769 |
| 243517 | 12222 | 12155 | 12248 | 12150 | 12104 | 12104 | 12150 | 12248 | 12155 | 12222 |
| 999917 | 49975 | 50129 | 50088 | 49883 | 49904 | 49904 | 49883 | 50088 | 50129 | 49975 |
| 2437 | 125 | 121 | 128 | 119 | 116 | 116 | 119 | 128 | 121 | 125 |
| 51637 | 2620 | 2542 | 2619 | 2583 | 2545 | 2545 | 2583 | 2619 | 2542 | 2620 |
| 92237 | 4611 | 4641 | 4640 | 4583 | 4584 | 4584 | 4583 | 4640 | 4641 | 4611 |
| 209837 | 10519 | 10496 | 10551 | 10460 | 10433 | 10433 | 10460 | 10551 | 10496 | 10519 |
| 997037 | 49840 | 49976 | 49953 | 49739 | 49751 | 49751 | 49739 | 49953 | 49976 | 49840 |
| 557 | 26 | 32 | 29 | 25 | 27 | 27 | 25 | 29 | 32 | 26 |
| 4157 | 208 | 214 | 215 | 201 | 201 | 201 | 201 | 215 | 214 | 208 |
| 33757 | 1721 | 1650 | 1717 | 1692 | 1659 | 1659 | 1692 | 1717 | 1650 | 1721 |
| 179957 | 8965 | 9090 | 9025 | 8938 | 8971 | 8971 | 8938 | 9025 | 9090 | 8965 |
| 977357 | 48926 | 48871 | 48988 | 48806 | 48748 | 48748 | 48806 | 48988 | 48871 | 48926 |
| 877 | 46 | 43 | 48 | 42 | 40 | 40 | 42 | 48 | 43 | 46 |
| 30677 | 1536 | 1549 | 1554 | 1516 | 1514 | 1514 | 1516 | 1554 | 1549 | 1536 |
| 15077 | 756 | 762 | 767 | 743 | 741 | 741 | 743 | 767 | 762 | 756 |
| 248477 | 12442 | 12449 | 12486 | 12380 | 12362 | 12362 | 12380 | 12486 | 12449 | 12442 |
| 985277 | 49247 | 49391 | 49358 | 49153 | 49170 | 49170 | 49153 | 49358 | 49391 | 49247 |
| 197 | 9 | 12 | 11 | 8 | 9 | 9 | 8 | 11 | 12 | 9 |
| 4597 | 238 | 224 | 241 | 227 | 219 | 219 | 227 | 241 | 224 | 238 |
| 18397 | 932 | 916 | 941 | 911 | 899 | 899 | 911 | 941 | 916 | 932 |
| 795997 | 39942 | 39653 | 39938 | 39804 | 39662 | 39662 | 39804 | 39938 | 39653 | 39942 |
| 989797 | 49567 | 49456 | 49611 | 49446 | 49369 | 49369 | 49446 | 49611 | 49456 | 49567 |

**Sequence when primes end in 9:**

In full length sequences, number of 0s is equal to the number of 9s which is one less than the frequency of any other digit. All other digits have the same frequency.

In a half length sequence that ends in 9, when the 2nd least significant digit is even, complementary digits are equal. It is also seen that digits 1, 2 and 4 and hence 5, 7 and 8 have the same frequency and are mostly minimum in number. 0s and 9s are mostly highest in number. In few of the sequences, 1s, 2s, 4s, 5s, 7s and 8s have highest frequency and 3s and 6s have the least frequency.

**Table 7:** Frequency distribution in the prime reciprocal sequences to the base-10 when the primes end in 9 and the 2nd least significant digit is even.

| Prime No | 0s | 1s | 2s | 3s | 4s | 5s | 6s | 7s | 8s | 9s |
|---|---|---|---|---|---|---|---|---|---|---|
| 409 | 24 | 19 | 19 | 21 | 19 | 19 | 21 | 19 | 19 | 24 |
| 3209 | 166 | 162 | 162 | 150 | 162 | 162 | 150 | 162 | 162 | 166 |
| 17609 | 891 | 892 | 892 | 835 | 892 | 892 | 835 | 892 | 892 | 891 |
| 194809 | 9832 | 9705 | 9705 | 9755 | 9705 | 9705 | 9755 | 9705 | 9705 | 9832 |
| 974009 | 48822 | 48750 | 48750 | 48430 | 48750 | 48750 | 48430 | 48750 | 48750 | 48822 |
| 929 | 50 | 47 | 47 | 41 | 47 | 47 | 41 | 47 | 47 | 50 |
| 4129 | 220 | 201 | 201 | 209 | 201 | 201 | 209 | 201 | 201 | 220 |
| 24329 | 1225 | 1226 | 1226 | 1179 | 1226 | 1226 | 1179 | 1226 | 1226 | 1225 |
| 254729 | 12790 | 12748 | 12748 | 12648 | 12748 | 12748 | 12648 | 12748 | 12748 | 12790 |
| 996529 | 49988 | 49766 | 49766 | 49846 | 49766 | 49766 | 49846 | 49766 | 49766 | 49988 |
| 1049 | 57 | 52 | 52 | 49 | 52 | 52 | 49 | 52 | 52 | 57 |
| 48449 | 2438 | 2441 | 2441 | 2351 | 2441 | 2441 | 2351 | 2441 | 2441 | 2438 |
| 56249 | 2817 | 2829 | 2829 | 2758 | 2829 | 2829 | 2758 | 2829 | 2829 | 2817 |
| 304849 | 15346 | 15196 | 15196 | 15278 | 15196 | 15196 | 15278 | 15196 | 15196 | 15346 |
| 996649 | 50044 | 49726 | 49726 | 49940 | 49726 | 49726 | 49940 | 49726 | 49726 | 50044 |
| 569 | 30 | 30 | 30 | 22 | 30 | 30 | 22 | 30 | 30 | 30 |
| 8969 | 459 | 452 | 452 | 427 | 452 | 452 | 427 | 452 | 452 | 459 |
| 46769 | 2347 | 2353 | 2353 | 2286 | 2353 | 2353 | 2286 | 2353 | 2353 | 2347 |
| 230369 | 11584 | 11523 | 11523 | 11439 | 11523 | 11523 | 11439 | 11523 | 11523 | 11584 |
| 993169 | 49833 | 49616 | 49616 | 49611 | 49616 | 49616 | 49611 | 49616 | 49616 | 49833 |
| 2089 | 113 | 101 | 101 | 106 | 101 | 101 | 106 | 101 | 101 | 113 |
| 30689 | 1543 | 1544 | 1544 | 1497 | 1544 | 1544 | 1497 | 1544 | 1544 | 1543 |
| 76289 | 3850 | 3820 | 3820 | 3762 | 3820 | 3820 | 3762 | 3820 | 3820 | 3850 |
| 602489 | 30248 | 30122 | 30122 | 30008 | 30122 | 30122 | 30008 | 30122 | 30122 | 30248 |
| 990889 | 49795 | 49447 | 49447 | 49586 | 49447 | 49447 | 49586 | 49447 | 49447 | 49795 |

In a half length sequence that ends with 9, if the second least significant digit is odd, sum of the frequencies of complementary digits for 0s is equal to integral part of $p/10$ and 1 more than integral part of $p/10$ for others. The digits 1, 5 and 6 have equal frequencies and hence the complementary digits 3, 4 and 8 have the same frequency. It is also seen that 0s or 2s are most frequent while 7s or 9s are least frequent in all such sequences.

As seen in case of 7 and 3, the digit frequency structure of half-length sequences that end in 9 and those that end in 1 have similar structure.

**Table 8:** Frequency distribution in the prime reciprocal sequences to the base-10 when the primes end in 9 and the 2nd least significant digit is odd.

| Prime No | 0s | 1s | 2s | 3s | 4s | 5s | 6s | 7s | 8s | 9s |
|---|---|---|---|---|---|---|---|---|---|---|
| 919 | 49 | 49 | 55 | 43 | 43 | 49 | 49 | 37 | 43 | 42 |
| 5519 | 316 | 280 | 288 | 272 | 272 | 280 | 280 | 264 | 272 | 235 |
| 23719 | 1209 | 1196 | 1216 | 1176 | 1176 | 1196 | 1196 | 1156 | 1176 | 1162 |
| 201119 | 10310 | 10078 | 10122 | 10034 | 10034 | 10078 | 10078 | 9990 | 10034 | 9801 |
| 994319 | 50080 | 49778 | 49902 | 49654 | 49654 | 49778 | 49778 | 49530 | 49654 | 49351 |
| 839 | 56 | 43 | 45 | 41 | 41 | 43 | 43 | 39 | 41 | 27 |
| 4639 | 243 | 239 | 253 | 225 | 225 | 239 | 239 | 211 | 225 | 220 |
| 34439 | 1812 | 1730 | 1746 | 1714 | 1714 | 1730 | 1730 | 1698 | 1714 | 1631 |
| 348239 | 17676 | 17436 | 17484 | 17388 | 17388 | 17436 | 17436 | 17340 | 17388 | 17147 |
| 994039 | 49752 | 49816 | 50044 | 49588 | 49588 | 49816 | 49816 | 49360 | 49588 | 49651 |
| 359 | 25 | 19 | 21 | 17 | 17 | 19 | 19 | 15 | 17 | 10 |
| 1759 | 93 | 92 | 100 | 84 | 84 | 92 | 92 | 76 | 84 | 82 |
| 23159 | 1233 | 1163 | 1173 | 1153 | 1153 | 1163 | 1163 | 1143 | 1153 | 1082 |
| 346559 | 17586 | 17354 | 17406 | 17302 | 17302 | 17354 | 17354 | 17250 | 17302 | 17069 |
| 999959 | 50546 | 50043 | 50133 | 49953 | 49953 | 50043 | 50043 | 49863 | 49953 | 49449 |
| 479 | 34 | 25 | 27 | 23 | 23 | 25 | 25 | 21 | 23 | 13 |
| 3079 | 164 | 159 | 169 | 149 | 149 | 159 | 159 | 139 | 149 | 143 |
| 44279 | 2304 | 2223 | 2241 | 2205 | 2205 | 2223 | 2223 | 2187 | 2205 | 2123 |
| 193679 | 10007 | 9703 | 9741 | 9665 | 9665 | 9703 | 9703 | 9627 | 9665 | 9360 |
| 961879 | 48255 | 48190 | 48382 | 47998 | 47998 | 48190 | 48190 | 47806 | 47998 | 47932 |
| 599 | 40 | 31 | 33 | 29 | 29 | 31 | 31 | 27 | 29 | 19 |
| 6199 | 307 | 321 | 343 | 299 | 299 | 321 | 321 | 277 | 299 | 312 |
| 41399 | 2172 | 2079 | 2097 | 2061 | 2061 | 2079 | 2079 | 2043 | 2061 | 1967 |
| 445799 | 22651 | 22323 | 22389 | 22257 | 22257 | 22323 | 22323 | 22191 | 22257 | 21928 |
| 989999 | 50242 | 49530 | 49590 | 49470 | 49470 | 49530 | 49530 | 49410 | 49470 | 48757 |

In addition to the structural redundancy seen in all these sequences, the prime numbers also are seen to have a definite structure. For each combination of the least significant digit and the next least significant digit, the third least significant digit is observed to be alternating between even and odd values when the second least significant digit is either even or odd. This property can be seen in every table above.

# Conclusions

We have shown several interesting structural properties of non-maximum length decimal sequences of prime reciprocals. We have classified these properties in relation to the digit in the least significant place of the prime. We have also observed, based on the experiment with primes less than a million, that the frequency of 0 exceeds that of other digits, in accordance with the corresponding result on binary prime reciprocal sequences.

If prime reciprocal sequences are used in cryptography applications, as has been proposed earlier, then these structural properties must be taken into consideration.